\documentclass{osa-article}

\usepackage{MnSymbol}

\journal{osac} 

\articletype{Research Article}

\begin{document}

\title{All-optical valley switch and clock of electronic dephasing}

\author{Rui E.F. Silva,\authormark{1} Misha Ivanov,\authormark{2,3,4} and \'Alvaro Jim\'enez-Gal\'an\authormark{2,5}}

\address{\authormark{1}ICMM, Centro Superior de Investigaciones Cient\'ificas, Madrid, Spain\\
\authormark{2}Max Born Institute, Max Born Strasse 2A, 12489 Berlin, Germany\\
\authormark{3}Department of Physics, Humboldt University, Berlin, Germany\\
\authormark{4}Blackett Laboratory, Imperial College London, London, United Kingdom\\
\authormark{5}Joint Attosecond Science Laboratory, National Research Council of Canada and University of Ottawa, Ottawa, Canada}

\email{\authormark{*}jimenez@mbi-berlin.de} 




\begin{abstract}
2D materials with broken inversion symmetry posses an extra degree of freedom, the valley pseudospin, that labels in which of the two energy-degenerate crystal momenta, K or K', the conducting carriers are located. It has been shown that shining circularly-polarized light allows to achieve close to $100\%$ of valley polarization, opening the way to valley-based transistors. Yet, switching of the valley polarization is still a key challenge for the practical implementation of such devices due to the short coherence lifetimes. Recent progress in ultrashort laser technology now allows to produce trains of attosecond pulses with controlled phase and polarization between the pulses. Taking advantage of such technology, we introduce a coherent control protocol to turn on, off and switch the valley polarization at faster timescales than electronic and valley decoherence, that is, an ultrafast optical valley switch. We theoretically demonstrate the protocol for hBN and MoS$_2$ monolayers calculated from first principles. Additionally, using two time-delayed linearly-polarized pulses with perpendicular polarization, we show that we can extract the electronic dephasing time $T_2$ from the valley Hall conductivity.
\end{abstract}



\section{Introduction}
\begin{figure}
\begin{center}
\includegraphics[width=\linewidth]{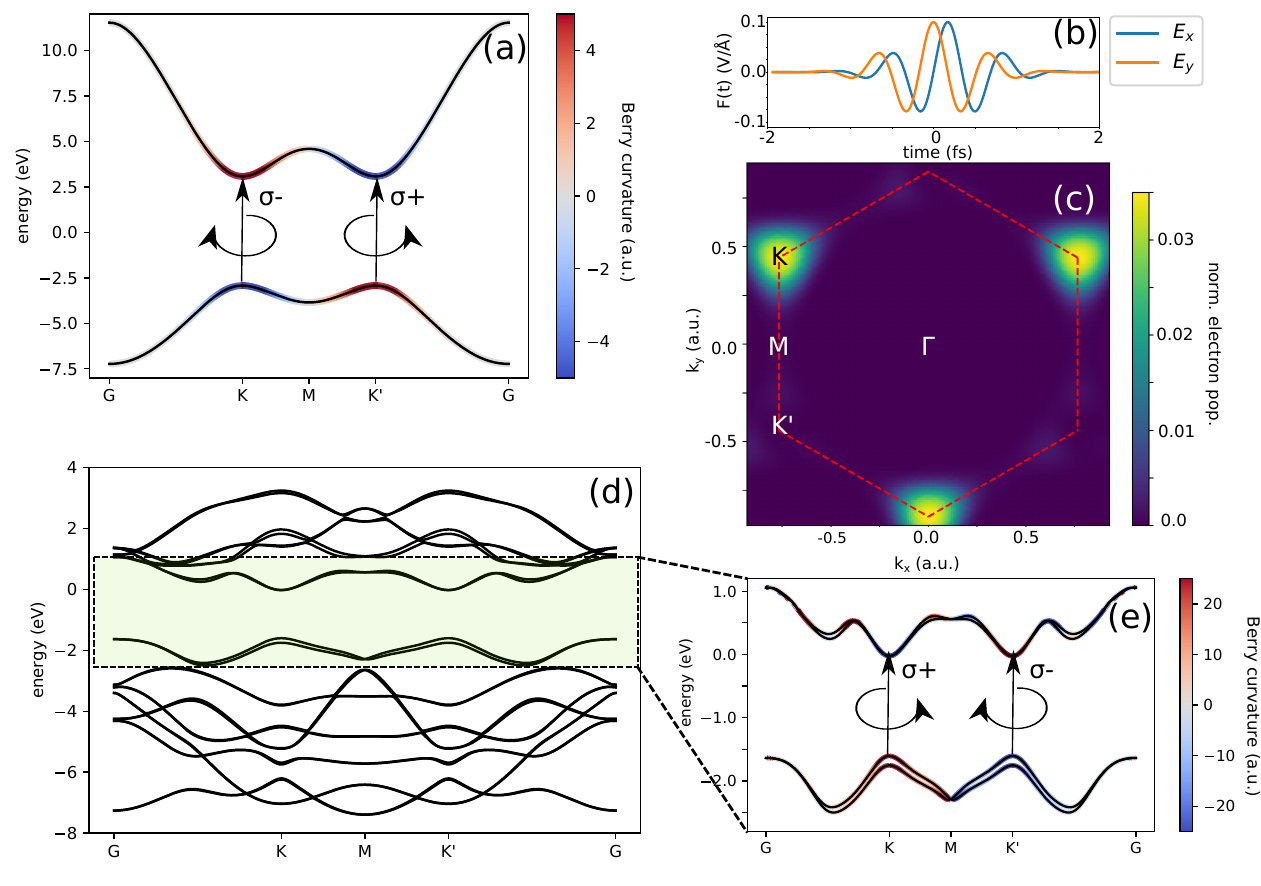}
\caption{\label{fig:bands} Band structure and optical valley selection rules. (a) Energy dispersion and Berry curvature (color map) for the $p_z$ bands in hBN. Arrows indicate the one-photon valley selection rules from each valley. (b) Few-cycle $\sigma_-$ circularly-polarized pulse. (c) Electron population in the first Brillouin zone (red-dashed hexagon) of the $p_z$ conduction band of hBN after the interaction with the pulse in (b). The population is normalized to 1 at each $k$-point. (d) Energy dispersion for MoS$_2$, including the $d$ orbitals of molybdenum and the $p$ orbitals of sulfur. (e) Expanded green region of panel (d), showing the relevant spin-split valence and conduction bands, their Berry curvature (color code) and the optical valley selection rules.}
\end{center}
\end{figure}

The synthesis of 2D materials with structure similar to graphene has opened the way to use and manipulate a new degree of freedom, the valley pseudospin, with potential applications in information processing and storage~\cite{Schaibley2016}. Valleys are energy minima of the electronic band structure of the crystal, which may be degenerate at different crystal momenta. In particular, for two-dimensional materials with a honeycomb structure and broken inversion symmetry, such as hexagonal boron nitride (hBN) or transition metal dichalcogenides (TMDs), the lowest conduction bands have energy-degenerate valleys located at the K and K' points of the Brillouin zone (see Fig.~\ref{fig:bands}a,d,e). The $C_3$ symmetry of the Bloch wavefunctions at the valleys leads to optical valley selection rules at $K$ and $K' \equiv -K$: $m_v(k) - m_c (k) \pm 1 = 3N$, where $N$ is an integer, the $\pm$ sign is determined by the the direction of rotation of the circular field, and the effective magnetic quantum numbers $m_{c,v} (k)$ can be related to the Berry curvature ~\cite{Xiao2010RevModPhys}.  

Fig.~\ref{fig:bands}a,e indicates the value of the Berry curvature of the relevant bands of hBN and MoS2, respectively, calculated from first principles~\cite{Jimenez-Galan:21}, along with the valley selection rules. The selection rule is opposite in hBN and MoS$_2$ due to the different orbital character of the bands. Moreover, the strong-spin orbit coupling in MoS$_2$ lifts the spin degeneracy at the valleys, allowing for a coupling between spin and valley pseudospin~\cite{Xiao2010RevModPhys}.
The light field can be only few-cycles long, as that in Fig.~\ref{fig:bands}b, allowing to induce valley polarization at few-femtosecond timescales (Fig.~\ref{fig:bands}c)~\cite{Motlagh2018}. 

One of the big obstacles for valleytronics are the short valley lifetimes, i.e., the time it takes the population to be scattered away from the valleys,  and the short coherence times, i.e., the phase coherence of a particle in a superposition of two valleys~\cite{Schaibley2016}. Previous works have estimated that the valley lifetime decays after 200~fs and the valley coherence after 100~fs for excitons in WSe$_2$ monolayers~\cite{hao2016direct}. Practical implementation of valleytronic devices therefore requires to induce valley polarization, switch it and read it at shorter timescales.

On the one hand, femtosecond control of the phase difference between the two valleys, i.e., valley coherence, has been achieved using the optical Stark effect~\cite{ye2017optical}. On the other hand, femtosecond valley polarization, i.e., inducing population at $K$ or $K'$, has been achieved using single-cycle pulses~\cite{Motlagh2018,Jimenez-Galan:21}, strong THz streaking fields~\cite{Langer:2018aa}, and strong tailored light fields~\cite{Jimenez2019}. In these methods, however, ultrafast switching of valley polarization, i.e., moving population from one valley to the other, is only possible while the material is dressed by the laser field. The high non-linearity of the process makes controlling the degree of valley polarization and its switching rate extremely challenging.

For a two-level system, population in the higher energy state can be switched on and off by two $\pi$-shifted resonant laser pulses delayed by $\tau = 2\pi/{\Delta E}$, where $\Delta E$ is the energy spacing between the two levels~\cite{warren1993coherent}. This is the simplest example of coherent control, which has seen a dramatic explosion thanks to the advancements in high intensity and ultrafast laser technology, which has allowed control and imaging of real-time motion of electrons in atoms, molecules and solids~\cite{baltuvska2003attosecond,gruson2016attosecond, kotur2016spectral, baltuvska2002controlling, lopez2005amplitude, dudovich2006measuring}. The production of trains of ultrashort pulses, down to the attosecond timescale, with a controllable phase and polarization relation between them is now possible~\cite{dorney2019controlling, pupeza2015high,Stummer:20}. Here, we theoretically demonstrate a coherent control protocol that uses a sequence of ultrashort pulses to completely control the valley polarization, i.e., turn it on and off as well as switch between the valley states, on timescales shorter than 100~fs. The valley polarization is sensitive to the time of electronic coherence between bands, additionally providing a tool to measure such times.

\section{Results}
Throughout the paper, we will be considering weak-field, resonant processes between the upper valence bands and lowest conduction bands of 2D materials. First, we illustrate our method in hBN described by the two $p_z$ bands closest to the band gap, as shown in Fig.~\ref{fig:bands}a. It will become clear that the arguments are general, and later we will apply the method to MoS$_2$ including 22 bands (Fig.~\ref{fig:bands}d,e). The light-matter interaction is simulated using the density matrix equations in the independent particle approximation using \emph{ab initio} equilibrium band structures and dipolar couplings, as described in previous works~\cite{Silva2019High,Jimenez-Galan:21}. An exponentially-decaying electronic dephasing with lifetime $T_2$ is included to account for the loss of coherence between the valence and conduction bands. We consider the particles to be electrons, and excitonic effects are not included due to the bigger numerical complexity. However, all of the results are equally applicable to excitons. In fact, the valley switching method we introduce here is presumed to work better with excitonic states, which are localized in the crystal momenta.

\begin{figure}
\begin{center}
\includegraphics[width=0.75\linewidth]{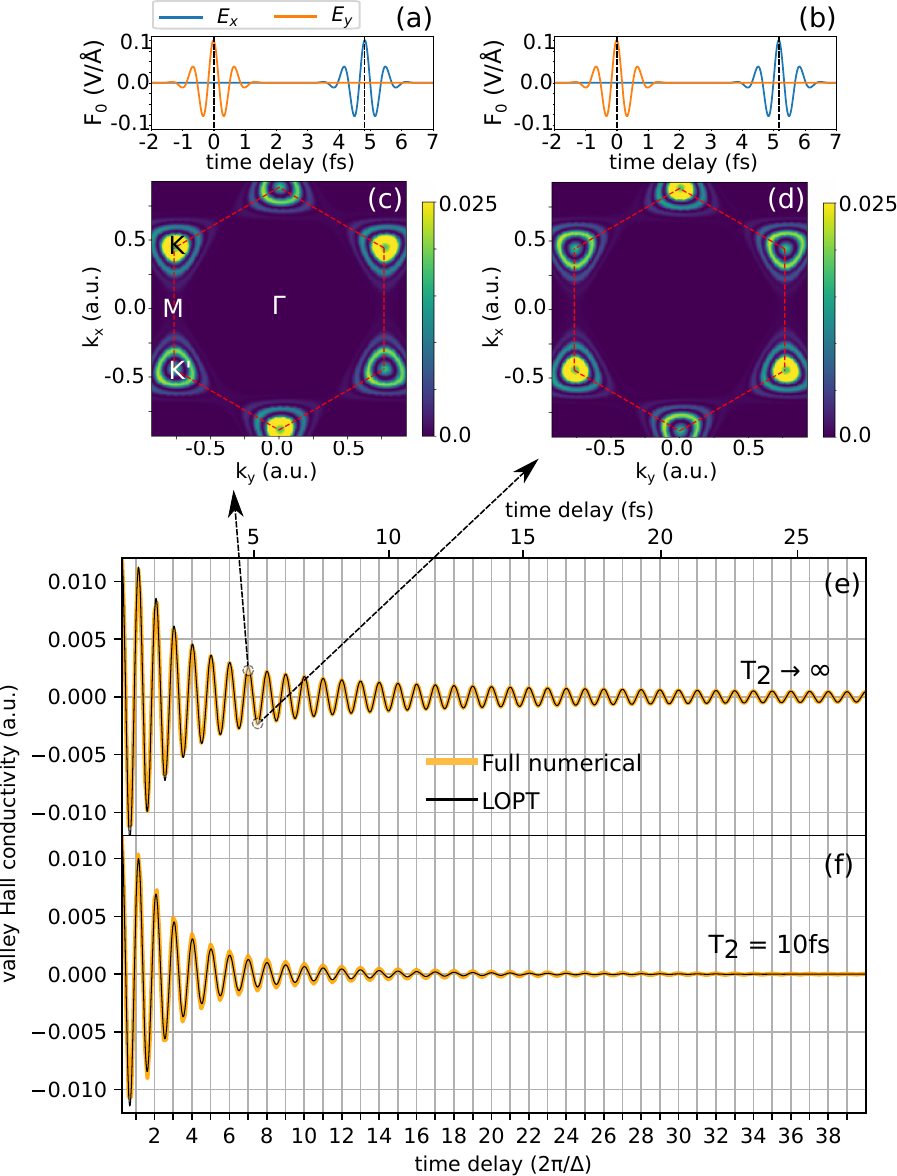}
\caption{\label{fig:dephasing} Measurement of electronic dephasing. (a,b) Sequence of two linearly-polarized pulses with perpendicular polarizations separated by a time delay of (a) $\tau = 4.88$~fs and (b) $\tau = 5.16$~fs. (c,d) Electron populations in the first Brillouin zone of the $p_z$ conduction of hBN for the pulses in (a) and (b), respectively. (e,f) Valley Hall conductivity as a function of the time delay between the two pulses, calculated for a dephasing time of (e) $T_2 \to \infty$ and (f) $T_2 = 10$~fs. The orange curves show the full time-dependent calculation while the black curves show the calculation using lowest order perturbation theory (LOPT). In (f), the LOPT calculation is multiplied additionally by a decaying exponential with lifetime $T_2=10$~fs (see text).
}
\end{center}
\end{figure}

To illustrate our method, let us first consider the problem where two linearly-polarized pulses with perpendicular polarizations and non-overlapping in time interact with hBN, as shown in Fig.~\ref{fig:dephasing}a,b. Besides their polarization direction, the two pulses in the sequence are identical. They are carried at the resonant band gap frequency $\omega=6$~eV, have a Gaussian envelope of 1.15~fs full width at half maximum and field strength of $F_0 = 0.1$~V/\AA. 

Fig.~\ref{fig:dephasing}c,d shows the electron populations in the conduction band which appear localized at $K$ and $K'$, respectively for the pulses in panels (a) and (b). The optical valley selection rules dictates that a linearly-polarized pulse does not induce valley polarization, while Fig.~\ref{fig:dephasing}c,d appears at first to show the opposite: the degree of valley polarization of the two linear, non-overlapping pulses is similar to that produced by a circularly-polarized field. This is possible thanks to the sustained electronic coherence between the two few-cycle pulses: any time delay $\tau = \frac{1}{\omega_{\mathcal{F}}} (\pm\pi/2 + 2\pi N)$, where $N$ is an integer, between a first $y$-polarized linear pulse and a second $x$-polarized linear pulse will produce a $\sigma_\mp$ circularly-polarized field at the frequency $\omega_\mathcal{F}$~\cite{warren1993coherent}. The dynamics at $\omega_\mathcal{F}$ will be equivalent for any chosen $N$ as long as the temporal separation of the pulses is smaller than the electronic coherence between the valence and conduction bands created by the first pulse, $\tau << T_2$. The valley polarization hence oscillates for different time delays between the two pulses (cf. panels c and d).

Due to the short duration of the pulses, the resulting bandwidth populates not only the $K$ and $K'$ points, but also many other nearby crystal momenta with different energies (see Fig.~\ref{fig:dephasing}c,d). Therefore, a time delay $\tau = \frac{1}{\Delta}\ (\pi/2 + 2\pi N)$, with $\Delta$ the band gap at $K$ and $K'$, does not guarantee maximum valley polarization. In order to obtain the optimal delay that yields the maximum valley polarization, a time delay scan must be performed first. Fig.~\ref{fig:dephasing}e,f shows the valley Hall conductivity (VHC) as a function of the time delay $\tau$, which is closely related to the valley polarization
\begin{equation}\label{eq:vhc}
\sigma (\tau) \propto \sum_n \int_{\text{BZ}} f_{n}(\mathbf{k},\tau) \Omega_{n} (\mathbf{k}) d\mathbf{k},
\end{equation}
where $n$ is the band index, $f_{n}(\mathbf{k},\tau)$ is the electron population at the $\mathbf{k}$ point after the two-pulse sequence ($t\to\infty$) for the time delay $\tau$, $\Omega_n (\mathbf{k})$ is the field-free Berry curvature, and the integral runs over the first Brillouin zone. Since $\Omega_n (\mathbf{k}) = -\Omega_n (-\mathbf{k})$, the VHC is zero for zero valley polarization, and changes sign for opposite valley polarization. Fig.~\ref{fig:dephasing}e,f reveals that in the region where the pulses are overlapping ($\tau < 2$~fs), the VHC peaks at $\tau = \frac{1}{\Delta}\ (\pm\pi/2 + 2\pi N)$, i.e., where the pulse is circular at the $K$ and $K'$ points. For longer delays, however, the VHC peaks at $\tau = \frac{\pi}{\Delta}\ N$, reflecting the dominant contributions from other crystal momenta. The valley polarization drops as a function of $\tau$ until it reaches a stable value for the case where there is no loss of electronic coherence, i.e., $T_2 \to \infty$ (panel e). These dynamics can be perfectly reproduced using lowest order perturbation theory (cf. orange and black curves in panel e), and are a consequence of the broadband wavepacket. The polarization phase of the pulse varies within the pulse bandwidth so that frequencies separated by $\Delta \omega$ differ by $\Delta \phi = \Delta \omega \tau$. As the time delay increases, the polarization of the field is varies more strongly between neighbouring frequencies, which will decrease the VHC. This decrease stabilizes at some point if $\tau << T_2$. However, for $\tau$ on the order of $T_2$, the loss of electronic coherence between the valence and conduction bands adds an additional decay mechanism of the VHC, as shown in Fig.~\ref{fig:dephasing}f, which completely cancels the signal for $\tau\,{\gtrapprox}\,2T_2$. This presents an opportunity to accurately retrieve the dephasing lifetime $T_2$.

Let us suppose we perform a time-delay measurement of the VHC on hBN and we obtain the orange curve in Fig.~\ref{fig:dephasing}f. For our purposes, we have simulated this curve from the full time-dependent calculation setting $T_2=10$~fs. To obtain the unknown $T_2$, we first need to obtain the reference function $\sigma_{\text{ref}}(\tau)$ that includes the drop of the VHC due to the wavepacket bandwidth but no electronic dephasing, which is given in Fig.~\ref{fig:dephasing}e. This is obtained numerically, either through a full-time dependent calculation (orange curve in panel e), or simply by using lowest-order perturbation theory (black curve in panel e). Once $\sigma_{\text{ref}}(\tau)$ is known, the lifetime $T_2$ can be extracted from the fit of our measured time-delay scan of the VHC (orange curve in panel f) to the function $\sigma(\tau,T_2) = \sigma_{\text{ref}}(\tau) e^{-\tau/T_2}$. Fig.~\ref{fig:dephasing}f shows the excellent agreement between the full time-dependent calculation using $T_2=10$~fs (orange curve) and $\sigma(\tau,T_2=10$~fs$)$ obtained through lowest order perturbation theory (black curve).

\begin{figure}
\begin{center}
\includegraphics[width=\linewidth]{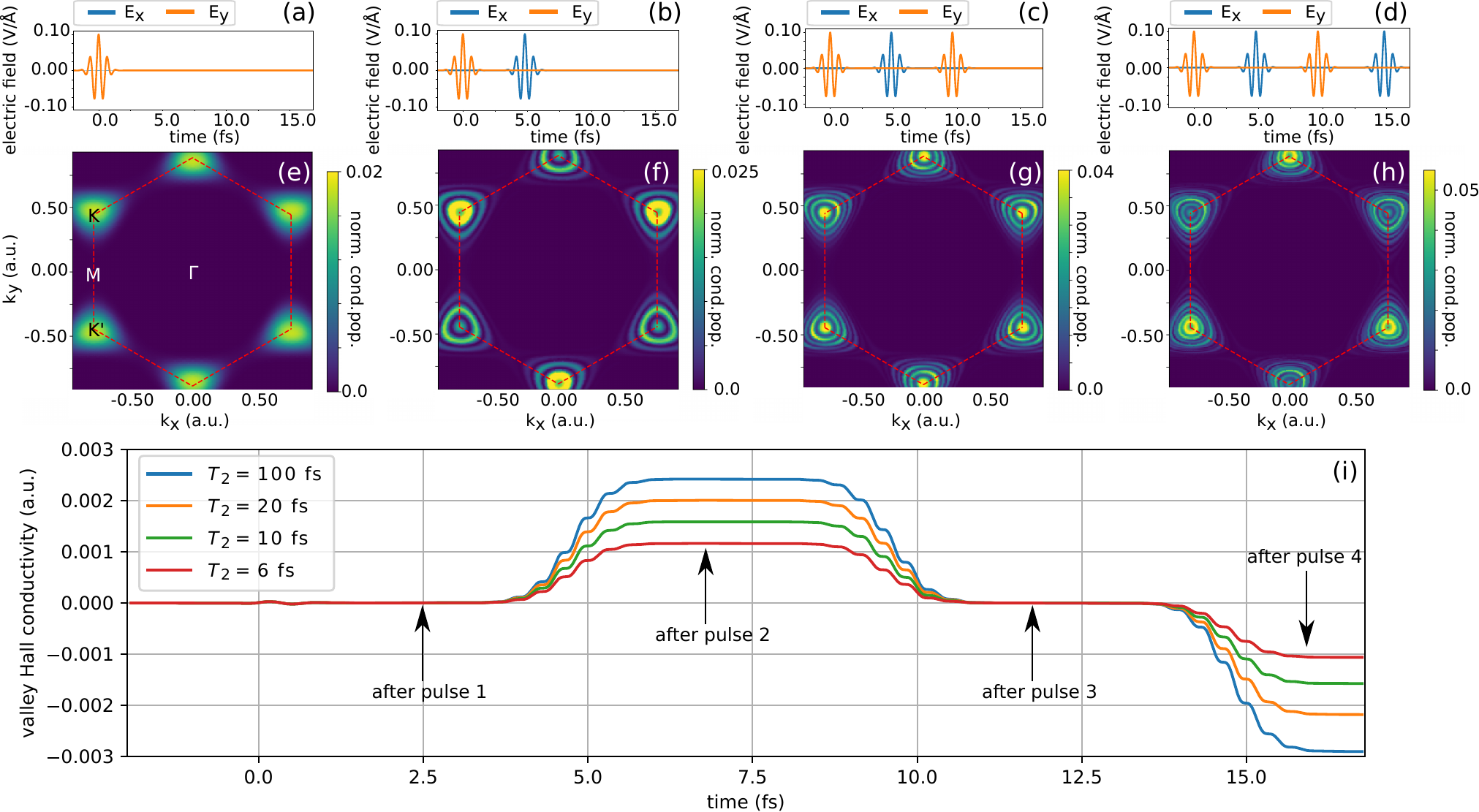}
\caption{\label{fig:hBN} Optical valley switch in a two band description of hBN. (a-d) Sequence of few-cycle pulses generating different values of the valley polarization. (e-h) Electron populations in the first Brillouin zone of the $p_z$ conduction band of hBN after the interaction with the pulses in panels (a-d), respectively. (i) Valley Hall conductivity as a function of time for the four-pulse sequence in panel (d), demonstrating the optical valley switch. Simulations were performed for different values of the dephasing time $T_2$, and are shown in different colors.
}
\end{center}
\end{figure}

Let us now use the concepts above to generate an optical valley switch. The protocol consists of a sequence of four identical linearly-polarized pulses resonant with the band gap, but polarized along different directions (Fig.~\ref{fig:hBN}a-d). First, we drive conduction band population using a field polarized along the y direction, although any other polarization direction is valid at this stage. The conduction band population in the Brillouin zone after the interaction with the pulse is shown in Fig.~\ref{fig:hBN}e. As expected from excitation by a linearly-polarized field, the electron population shows no valley polarization. This first pulse creates coherence between the valence and conduction bands. 

At an appropriate later time $\tau$, obtained through a time-delay scan of the VHC $\sigma (\tau)$, we make the system interact with a second linearly-polarized pulse polarized perpendicular to it (Fig.~\ref{fig:hBN}b). As we discussed earlier, the maximum valley polarization at $K$ is obtained for $\tau = N\times 2\pi/\Delta$, and we choose $\tau = 7 \times 2\pi/\Delta = 4.8$~fs, which guarantees both a reasonably high value of valley polarization and non-overlapping pulses. The valley polarization at $K$ is clear (Fig.~\ref{fig:hBN}f). Until now, we have simply induced valley polarization, no more than what we would have achieved with a single circularly-polarized pulse. Let us now switch off this valley polarization. 

For this, we add a third pulse, linearly-polarized along the same direction of the first, as shown in Fig.~\ref{fig:hBN}c. In this way, we are able to choose a time delay for the third pulse which completely switches off the valley polarization (Fig.~\ref{fig:hBN}g). The time delay chosen is $\tau = 14 \times 2\pi/\Delta = 9.6$~fs, again obtained through a time-delay scan of the VHC. We have now managed to switch on and off the valley polarization on a timescale of 10~fs, one order of magnitude shorter than the estimated times of valley decoherence~\cite{hao2016direct} and electronic dephasing~\cite{kilen2020propagation}.

To finalize the scheme, we seek to induce the opposite valley polarization as that induced by the second pulse. We add a fourth pulse linearly-polarized along the same direction as the second pulse (Fig.~\ref{fig:hBN}d). In this case, the optimal time delay for this operation is $\tau = 21.5 \times 2\pi/\Delta = 14.8$~fs. Fig.~\ref{fig:hBN}h shows the switch of the valley polarization with respect to that induced by the second pulse. All of the protocol is performed on timescales one order of magnitude shorter than decoherence times.

Fig.~\ref{fig:hBN}i shows the time-dependent VHC, $\sigma (t)$, for the four-pulse sequence in Fig.~\ref{fig:hBN}d calculated using electronic dephasing times ranging from $T_2=100$~fs, which is the estimated order of magnitude reported in recent work~\cite{kilen2020propagation,Floss2018}, to more than one order of magnitude shorter. Note that $\sigma(t)$ is slightly different to $\sigma(\tau)$ defined in Eq.~\ref{eq:vhc}, since now $f_n (\mathbf{k},t)$ refers to the population during the interaction with the four-pulse sequence (from $t=-2$~fs to $t=17.5$~fs). The VHC at the end of the scheme ($t \simeq 17.5$) drops by about one third between these two dephasing times, but the valley switching is still clearly visible even for a separation between the first and last pulse of $\tau = 2.5\,T_2$, demonstrating the robustness of our method.

The method described above is general. While the large band gap of hBN allows to use resonant few-cycle pulses with very short duration, it also makes hBN challenging to be used as an optically-pumped valleytronic material. On the other hand, monolayer TMDs are perfect candidates for this end, since their direct band gaps lie in the visible range (between 400~nm and 700~nm)~\cite{Schaibley2016}. This allows to use conventional Ti:Sapphire lasers to induce valley polarization. TMDs also display strong dipole couplings, most prominently to excitonic states. Here, we will not consider excitonic transitons, but instead the resonant electronic transition between the valence bands and the conduction bands, as we did with hBN, and as shown in Fig.~\ref{fig:bands}e. We note, however, that the protocol would be equivalent for excitonic states, with the additional advantage that the wavepacket will be more localized in $k$-space and therefore in energy. This means that the decay of the valley polarization of electrons in the conduction band discussed in Fig.~\ref{fig:dephasing}e will be less pronounced in the case of excitonic states.

\begin{figure}[h!]
\begin{center}
\includegraphics[width=\linewidth]{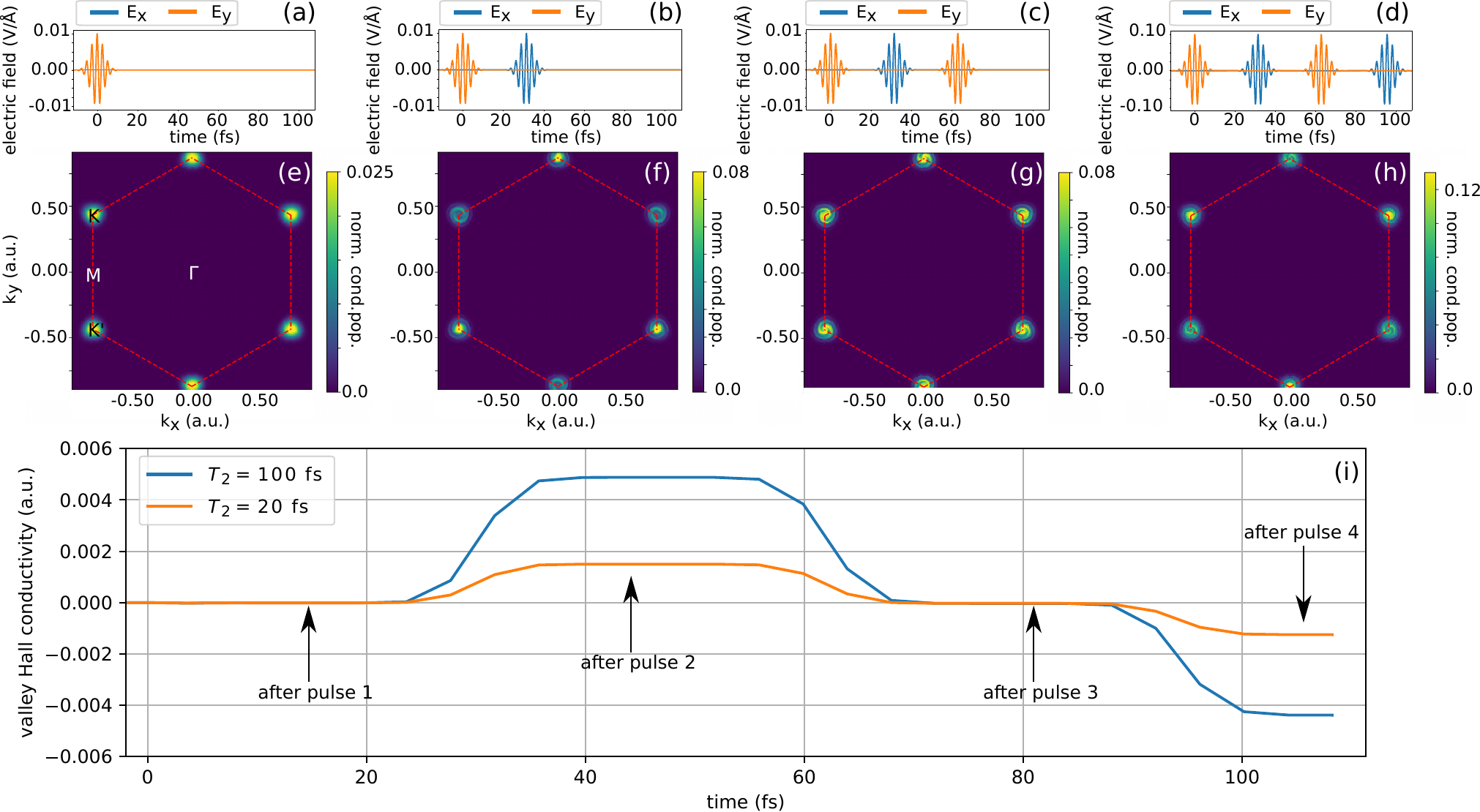}
\caption{\label{fig:MoS2} Optical valley switch in a 22 band description of MoS$_2$. (a-d) Pulse sequences generating different values of the valley polarization. (e-h) Sum of the electron populations in the first two conduction bands of MoS$_2$ (see Fig.~\ref{fig:bands}e) after the interaction with the pulses in panels (a-d), respectively. (i) Valley Hall conductivity as a function of time for the four pulse sequence in panel (d), computed from the sum of the electron populations in the two lowest conduction bands. Results for dephasing times $T_2=100$~fs and $T_2=20$~fs are shown in blue and orange, respectively.}
\end{center}
\end{figure}

Fig.~\ref{fig:MoS2} shows the protocol in MoS$_2$, where we included the 22 bands shown in Fig.~\ref{fig:bands}d in the time-dependent simulations. The pulse parameters, shown in Fig.~\ref{fig:MoS2}a-d, are in this case: $\omega = 1.58$~eV, matching the ab-initio minimum band gap of MoS$_2$ (see Fig.~\ref{fig:bands}d,e), field strength $F_0 = 0.01$~V/\AA, and Gaussian pulse envelope with $6.8$~fs full width at half maximum. Fig.~\ref{fig:MoS2}e-h shows that the valley polarization is turned on, off and switched analogously to hBN, apart from a sign change due to the opposite valley selection rules (cf. Fig.~\ref{fig:bands}a,e). Fig.~\ref{fig:MoS2}i reveals that the VHC drops by a factor of 4 between $T_2=100$~fs and $T_2=20$~fs. The switch of the sign in the VHC is nonetheless still clearly visible for $T_2=20$~fs, which is one order of magnitude smaller than the predicted electronic dephasing time. Both the electron populations in panels (e)-(h) and the valley Hall conductivity in panel (i) were computed by adding up the populations of the two lower spin-split conduction bands (see Fig.~\ref{fig:bands}e), where 99$\%$ of the excited population resided.

\section{Conclusion}
In summary, by using a sequence of ultrashort pulses with controlled phase delay and polarization, we have demonstrated an all-optical protocol to turn on, off and switch the valley polarization on timescales shorter than reported decoherence times, and well within experimental reach. We have applied our method to hBN, described by two bands, and MoS$_2$, described by 22 bands, calculated from first principles, obtaining equally good results.  We have also shown that by measuring the valley Hall conductivity as a function of the time delay between two perpendicularly-polarized pulses, the electronic dephasing time $T_2$ can be retrieved. Our work gives important steps towards the practical implementation of ultrafast all-optical valley switches and towards the determination of electronic coherence times in 2D materials.

$ $

\noindent\textbf{Funding.} R.E.F.S. acknowledges support from the fellowship LCF/BQ/PR21/11840008 from `La Caixa` Foundation (ID 100010434). \'A.J.-G. acknowledges funding from the European Union Horizon 2020 research and innovation programme under the grant agreement no. 101028938.

$ $

\noindent\textbf{Disclosures.} The authors declare no conflicts of interest.

\bibliography{biblio}



\end{document}